\documentstyle[11pt,newpasp,twoside]{article} 
\def\edcomment#1{\iffalse\marginpar{\raggedright\sl#1\/}\else\relax\fi} 
\reversemarginpar 

\begin{document} 
\title{Globular Cluster Populations: Signatures and Implications}

\author{Uta Fritze--v. Alvensleben$^{(1)}$, Richard de Grijs$^{(2)}$} 
\affil{$^1$ Universit\"ats-Sternwarte, Geismarlandstr. 11, 37083 G\"ottingen, Germany\\ 
$^2$ Institute of Astronomy, Madingley Road, Cambridge, CB3 0HA, UK}

\section{Introduction} 
The formation of populous secondary star cluster systems is a widespread phenomenon in mergers of gas-rich galaxies. Many, if not most, of those clusters are massive and compact enough to be young globular clusters (GCs). GC systems in most E/S0 galaxies feature bimodal color distributions with a fairly universal blue peak similar to the blue peak of halo GCs in the Milky Way (MW) and M31, and a variable red peak. Due to the well-known age -- metallicity degeneracy of optical broad-band colors, the metallicities and ages, and, hence, the origin of the red peak GCs are not yet known. 
We use evolutionary synthesis models for GC {\bf systems} of various metallicities to study the time evolution of their luminosity functions (LFs) in various bands U,..., K and of their color distributions. By comparison with the universal blue peak GC population we investigate for which combinations of age and metallicity a second GC population can or cannot be identified in typical observations of GC color distributions and we discuss implications for the GC LF as a distance indicator.

\section{Evolutionary Synthesis Models for GC Populations}
Our evolutionary synthesis models for {\bf S}imple {\bf S}tellar {\bf P}opulations (SSPs $=$ single burst single metallicity), based on isochrones from the Padova group, combined with stellar model atmosphere libraries for different metallicities extend over wavelengths from the UV through the NIR (cf. Schulz et al. 2002). They provide theoretical color -- metallicity calibrations which for old and low metallicity GCs agree well with empirical calibrations from MW or M31 GCs. However, they also show that the theoretical calibrations steepen considerably for metallicities [Fe/H]$>-0.5$ and change significantly for ages younger than $\sim 11$ Gyr. Hence, at a given color, the theoretical calibrations predict considerably lower metallicities for clusters with [Fe/H]$>-0.5$ and considerably higher metallicity for clusters younger than 11 Gyr. We stress that -- as long as ages and metallicities of the red-peak GCs are not known -- empirical color -- metallicity calibrations should not be used for their analysis. 

The universal blue peak in bimodal color distributions of E/S0 GC systems is compatible with SSP models for [Fe/H]$= -1.6$ and age $\sim 12$ Gyr. We take this universal blue GC population as a reference for a typical primary GC population and assume that the width of the color distribution is the same for the secondary and primary GC populations. We then analyse for what combinations of age and metallicity a second GC population might be distinguished observationally from a primary one and which combinations might explain the red peak in bimodal optical GC color distributions. 

For sample sizes of 100 -- 200 (500 -- 1000) GCs two peaks can be distinguished observationally if they are separated by at least 2.5 -- 3 (2) times the observational uncertainty in the respective color (Ashman, Bird \& Zepf 1994). Our models show that for many combinations of ages and metallicities a second GC population may be hidden within the blue peak in optical colors like ${\rm V-I}$, but should clearly be detectable in optical$-$NIR or UV$-$optical colors, and that many combinations of age and metallicity are possible for the optical red-peak of GC populations, which, in turn, should result in significant differences in other colors (cf. Fritze -- v. Alvensleben 2002, A\&A {\sl in prep.}). We show that multi-color observations including the NIR (and ideally also the UV) will allow to disentangle ages and metallicities of the GC population(s) in a red peak and, hence, shed light on the time and environment of their formation. 

\section{Effects of Bimodal GC Color Distributions on LFs and H$_0$}
For all the metallicities we also explore the resulting LFs {\bf in their time evolution} assuming that the width of the LF is dominated by the width of the star cluster mass function also assumed to be the same for both GC populations, and we compare to the LF of the blue-peak GCs. As an example, we here apply our models to the GC system in NGC 4472. Larsen et al. (2001) show that it has a bimodal color distribution in ${\rm V-I}$ with peaks at ${\rm V-I=0.943}$ and 1.207, respectively, and they also determine the LFs separately for the blue- and red-peak GCs, ${\rm \langle M_V \rangle_b= -7.55}$ and ${\rm \langle M_V \rangle_r= -7.05}$ mag. We find the red peak to allow for a wide range of age -- metallicity combinations ([Fe/H], age) ${\rm = (-0.4,\;14~Gyr),~(-0.2,\;10~Gyr),~(0,\;6~Gyr),~(+0.4,\;2.5~Gyr)}$ and predict ${\rm (V-K,\, \langle M_V \rangle)}$ values as different as ${\rm (3.10,\,-6.7),~(3.23,\,-6.9)}$, ${\rm (3.30,\,-7.1)}$, ${\rm (3.60,\,-7.7)}$ for these. Additional K-band observations allow to select one of these combinations and the comparison of the observed red-peak ${\rm \langle M_V \rangle_r}$ with the one predicted by our models under the assumption of a uniform underlying cluster mass function will allow to verify or reject this assumption conclusively. It does not seem to make sense in systems with bimodal color distributions to use ``the'' GC LF as a distance estimator and for the determination of the Hubble constant. Note that the observed difference of 0.5 mag between the red and blue GC LFs in NGC 4472 translates into a difference ${\rm \Delta H_0/H_0 \ga 15 \%}$! If the blue peak of bimodal GC color distributions will be confirmed to be as universal as it presently seems, we suggest that the use of the LF of the blue peak clusters may serve as a very precise distance indicator out to large distances $\ga 100$ Mpc and allow for reliable estimates of H$_0$.

\end{document}